\documentclass[fleqn,usenatbib]{mnras}

\usepackage{newtxtext,newtxmath}
\usepackage[T1]{fontenc}

\DeclareRobustCommand{\VAN}[3]{#2}
\let\VANthebibliography\thebibliography
\def\thebibliography{\DeclareRobustCommand{\VAN}[3]{##3}\VANthebibliography}

\usepackage{graphicx}	% Including figure files
\usepackage{amsmath}	% Advanced maths commands
\usepackage{ulem}

\title[Binaries in Omega Centauri]{Studying Binary Systems in Omega Centauri with MUSE. I. Detection of Spectroscopic Binaries}
\author[F.Wragg et al.]{
F. Wragg,$^{1}$\thanks{E-mail: F.E.Wragg@2020.ljmu.ac.uk}
S. Kamann,$^{1}$\thanks{E-mail: s.kamann@ljmu.ac.uk}
S. Saracino,$^{1}$
M. Latour,$^{2}$
S. Dreizler,$^{2}$
S. Martens,$^{2}$
A. Seth,$^{3}$
\newauthor
D. Vaz,$^{4, 6}$
G. van de Ven,$^{5}$
\\
$^{1}$Astrophysics Research Institute, Liverpool John Moores University, 146 Brownlow Hill, L3 5RF, UK\\
$^{2}$Institut f\"{u}r Astrophysik und Geophysik, Georg-August-Universit\"{a}t G\"{o}ttingen,, Friedrich-Hund-Platz 1, 37077, Germany\\
$^{3}$Department of Physics and Astronomy, University of Utah, Salt Lake City, UT 84112, USA\\
$^{4}$Instituto de Astrof\'{i}sica e Ci\^{e}ncias do Espa\c{c}o, Universidade do Porto, Rua das Estrelas, 4150-762 Porto, Portugal\\
$^{5}$Department of Astrophysics, University of Vienna, T\"{u}rkenschanzstrasse 17, A-1180 Vienna, Austria\\
$^{6}$Departamento de F\'{i}sica e Astronomia, Faculdade de Ci\^{e}ncias, Universidade do Porto, Rua do Campo Alegre 687, PT4169-007 Porto, Portugal
}

\date{Accepted XXX. Received YYY; in original form ZZZ}

\pubyear{2023}

\begin{document}
\label{firstpage}
\pagerange{\pageref{firstpage}--\pageref{lastpage}}
\maketitle

\begin{abstract}

NGC 5139 ($\omega$ Cen), is the closest candidate of a Nuclear Star Cluster that has been stripped of its host galaxy in the Milky Way. Despite extensive studies through the last decades, many open questions about the cluster remain, including the properties of the binary population. In this study we use MUSE multi-epoch spectroscopy to identify binary systems in $\omega$ Cen. The observations span 8 years, with a total of 312\,248 radial velocity measurements for 37\,225 stars. Following the removal of known photometric variables, we identify 275 stars that show RV variations, corresponding to a discovery fraction of $1.4\pm0.1\%$. Using dedicated simulations, we find that our data is sensitive to $70 \pm 10\%$ of the binaries expected in the sample, resulting in a completeness-corrected binary fraction of $2.1\pm0.4\%$ in the central region of $\omega$ Cen. We find similar binary fractions for all stellar evolutionary stages covered by our data, the only notable exception being the blue straggler stars, which show an enhanced binary fraction. We also find no distinct correlation with distance from the cluster centre, indicating a limited amount of mass segregation within the half-light radius of $\omega$ Cen.

\end{abstract}


%%%%%%%%%%%%%%%%%%%% MAIN TEXT %%%%%%%%%%%%%%%%%%

\section{Introduction}
    \label{Sec:Intro}    

    Binary systems are a key element to the evolution of stellar clusters. They act as both a source and store of gravitational energy, which can delay core collapse in globular clusters \citep{Goodman_1989,Breen_Heggie_2012}. Binary stars are also vital to explain the variety of stellar exotica observed in star clusters, such as blue stragglers \citep[BSS, e.g.,][]{McCrea_1964,Mateo_1990,Stryker_1993,Giesers_2019} and cataclysmic variables \citep[CV, e.g.,][]{Ivanova_2006,Goettgens_2019}. Finally, binary systems composed of stellar remnants and luminous companions offer a unique possibility to constrain the populations of neutron stars and black holes residing in star clusters. Doing so is crucial in order to enhance our understanding of the origin of gravitational wave (GW) events detected by the LIGO-VIRGO-KAGRA collaboration.

    Black holes have been suggested to exist in globular cluster environments for a long time, however it was initially thought that only a small number would survive the natal kicks experienced by forming black holes \citep{Kulkarni_1993, Sigurdsson_1993} and any black holes that survived would become an isolated subsystem at the centre of the cluster due to the \citet{Spitzer_1969} instability. Recent observations of black hole candidates in globular clusters \citep{Strader_2012, Minniti_2015, Giesers_2018}, suggest that black holes are neither exceptional objects nor isolated from luminous cluster members. The longevity of black holes in star clusters is also predicted by state-of-the-art numerical models, such as Monte Carlo or N-body models \citep[e.g.,][]{Askar_2018,Weatherford_2020}. Black hole mergers in globular clusters are thought to be an important source of GWs, with subsequent simulations such as those performed by \cite{Rodriguez_2016} and \cite{Antonini_2020} predict a relatively high merger rate in globular cluster environments.
    
    Although among field stars the binary fraction is $>50\%$ \citep{Duquennoy_1991, Duchene_2013}, globular clusters typically have a lower binary fraction, with the spectroscopic studies of M4 by \cite{Sommariva_2009} and of NGC 3201 by \cite{Giesers_2019} reporting global binary fractions of $3.0\%\pm0.3\%$ and $6.75\%\pm0.72\%$, respectively. The photometric study of 59 galactic globular clusters by \cite{Milone_2012_photometry} also supports a lower global binary fraction, which for most clusters falls between $2\%$ and $15\%$. The discrepancy compared to the field can be explained by the large frequency of dynamical interactions within star cluster environments, which tend to destroy wide binaries \citep{Heggie_1975}. On the other hand, the work by \citet{Heggie_1975} further shows that dynamical interactions tend to harden tight binaries, and indeed, binary interaction products, such as BSS, are commonly found in dense clusters. As a consequence, the observed binary fractions cannot be easily converted to primordial ones \citep{Hut_1992}, and the latter are still poorly constrained. Simulations of globular clusters using widely ranging primordial fractions between 5\% \citep{Hurley_2007} and 100\% \citep{Ivanova_2005} are able to reproduce the observed fractions.

    Studies of young massive clusters (ages $\lesssim$ 10Myr), such as 30 Doradus \citep{Sana_2013, Dunstall_2015}, Westerlund~1 \citep{Ritchie_2022}, or NGC 6231 \citep{Banyard_2022} report binary fractions between $40$ and $55\%$. These results suggest, assuming young massive clusters do in fact represent the early stages of globular cluster evolution, that the primordial binary fractions of globular clusters were indeed much higher than the fractions observed today. However, care must be taken when comparing young massive clusters and ancient globulars, given that the results available for the two classes of objects are derived for very different stellar mass regimes and that the multiplicity properties of field stars are a strong function of stellar mass \citep[see][for a recent overview]{Offner_2023}.
    
    Within globular clusters, the binary fraction is seen to decrease with radius \citep[e.g.,][]{Milone_2012}, which can be understood as a consequence of mass segregration \citep{Hurley_2007, Fregeau_2009}. As shown by, e.g., \citet{Aros_2021}, the amount of mass segregation experienced by binary stars sensitively depends on the number of black holes, or the presence of an intermediate-mass black hole (IMBH), inside the cluster. Namely, less mass segregation corresponds to the presence of a substantial black hole component, either as a higher number of stellar mass black holes, or an IMBH. As such, studying the properties of binary systems inside globular clusters promises to offer new insight into the distribution of unseen mass inside the clusters. 

%It is an unusually complex cluster with up 17 populations of stars with distinct chemistry \citep{Bellini_2017}

    NGC 5139, Omega Centauri ($\omega$ Cen), is a particularly important and interesting cluster in the Milky Way. Due to its unusual complexity, both in terms of the chemical abundances of stars \citep{Alvarez_2024}, but also in the clusters kinematics and internal dynamics \citep{van_de_Ven_2006}, $\omega$ Cen has been suggested a remnant of a dwarf galaxy that has merged with the Milky Way and is proposed to be a former Nuclear Star Cluster (NSC), though there is still no consensus on whether $\omega$~Cen belonged to either the Gaia-Enceladus or the Sequoia dwarf galaxies \citep{Massari_2019, Forbes_2020, Pfeffer_2021}. Determining the binary fraction of a cluster as complex as $\omega$ Cen will provide a crucial insight into the evolution of the cluster. Given the potential link between $\omega$~Cen and a former dwarf galaxy, there is a long-standing question about the presence of an IMBH in the cluster. However, until now, uncertainties regarding the location of the cluster centre \citep{Anderson_2010} or the retention fraction of stellar-mass black holes have prevented a clear answer \citep[see debate in][]{vanderMarel_2010,Noyola_2010}. N-body models performed by \citet{Baumgardt_2019} suggest that a large number of stellar-mass black holes must have been retained in order to explain the observed stellar kinematics of $\omega$~Cen. A spectroscopic binary search offers the possibility to partially uncover this predicted population, in a similar way as done by \citet{Giesers_2018, Giesers_2019} for NGC~3201. At the same time, the radial distribution of binaries promises further insight into the mass structure of $\omega$~Cen, as predicted by \citet{Aros_2021}. There have been very few direct investigations for binary systems in $\omega$ Cen, and no studies that give an accurate value for the global fraction of the cluster. The most recently published estimates for binary fraction are photometric studies by \cite{Bellini_2017}, which estimates a fraction of $2.70 \pm 0.08\%$, and one by \cite{Elson_1995}, which estimates an upper limit of $5\%$.

    In this first of its kind study of $\omega$ Cen, we conduct a spectroscopic search for binary systems within the cluster. In Section \ref{Sec:Data}, we discuss the spectroscopic data collected using the Multi Unit Spectroscopic Explorer (MUSE) and describe the statistics of the observations and the data reduction. In Section \ref{Sec:Analysis}, we summarize the process of spectral extraction and analysis used to perform the radial velocity measurements for each epoch of each star. In Section \ref{Sec:Variability_Meaurement}, we remove variable stars from the sample, and calculate the discovery fraction of radial-velocity variables in our data set of $\omega$ Cen. We describe in Section \ref{Sec:Completeness_Correction} the procedure used to determine the completeness of our binary search. Finally, in Section \ref{Sec:Results}, we discuss the robustness of the mock data sample and the sensitivity to the assumed orbital parameter distributions.  We then present the completeness-corrected binary fraction, not only for the global population, but also for different areas of the CMD, before summarising our results in Section \ref{Sec:Conclusions}.

\section{Data}
    \label{Sec:Data}
    
    \subsection{Observations}
        \label{Subsec:Observations}
        
        One of the biggest challenges in spectroscopic observations of globular clusters is the crowding of the field, resulting in the blending of sources. As a consequence, spectroscopic samples large enough to determine the demographics of the binary populations in clusters like $\omega$~Cen are still scarce. However, the Multi Unit Spectroscopic Explorer \citep[MUSE,][]{Bacon_2010} at the Very Large Telescope (VLT) saw first light in 2014 and since has observed 27 globular clusters within the Milky Way, including $\omega$ Cen, as part of the guaranteed time observations \citep[GTO, see][for an overview of the observations]{Kamann_2018}. In the crowded cluster fields, MUSE enables simultaneous spectroscopy of thousands of stars.
        
\begin{figure}
            \centering
            \includegraphics[scale=0.36]{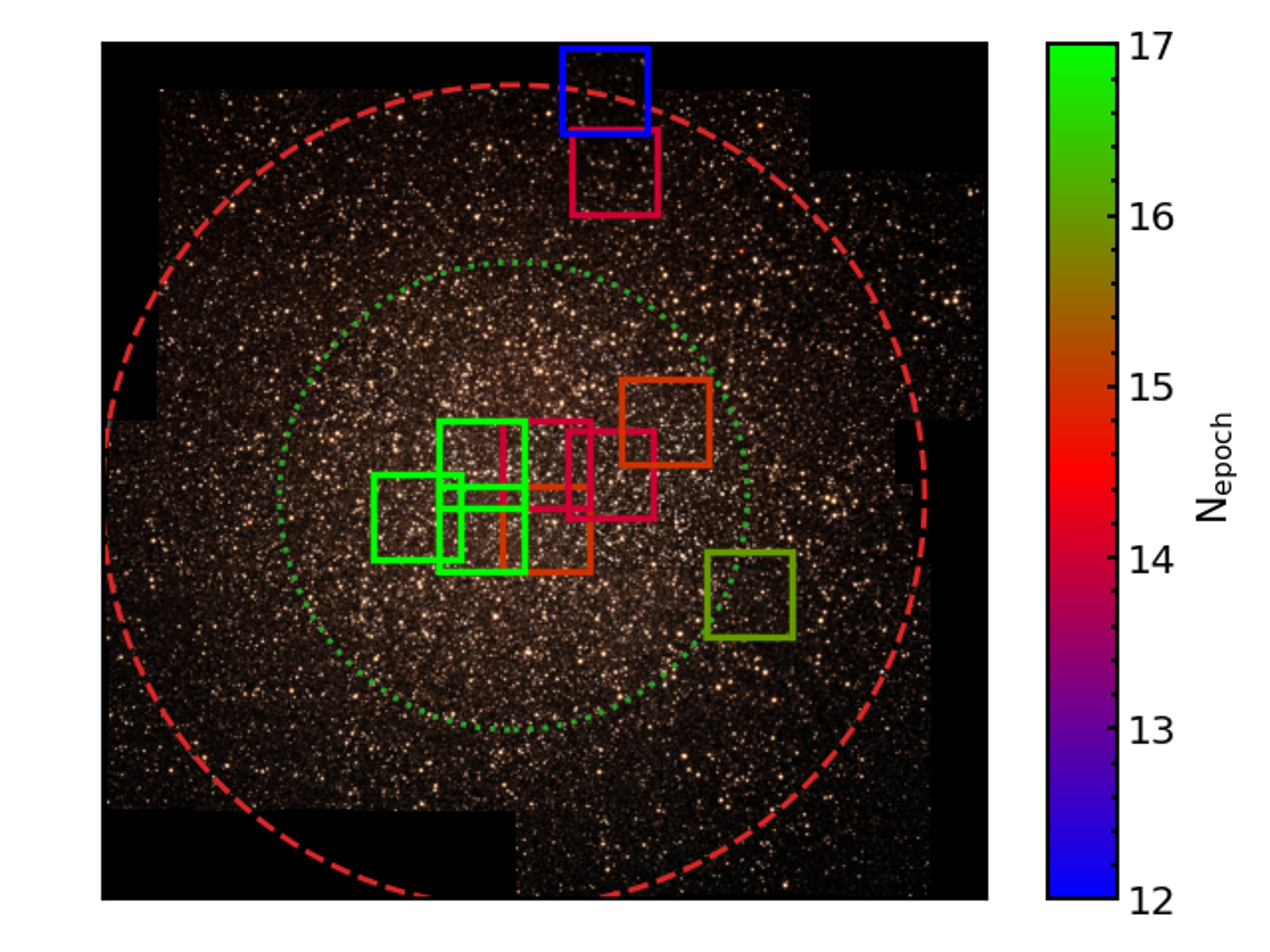}
            \caption{VRI colour representation of the MUSE data presented in \mbox{\citet{Nitschai_2023}}, with the GTO pointings studied in this work highlighted, with the number of epochs for each pointing shown via the colourbar. The 6 NFM (7.5" across) pointings are not shown as they would be too small to see, but are in the overlap region of the four central WFM pointings. The green dotted circle represents the clusters core radius, and the red dashed circle shows the half light radius. In this image, up is north and left is east.}
            \label{fig:MUSE_Mosaic}
        \end{figure}

        For this study, we use all the MUSE GTO data available of $\omega$ Cen: 10 wide field mode (WFM) pointings with a 1' $\times$ 1' field of view (FoV), and 6 narrow field mode (NFM) pointings with a 7.5" $\times$ 7.5" FoV were observed repeatedly between 2015 and 2022. For a detailed summary of the individual pointings observed in $\omega$~Cen, see the recent work of \citet{Pechetti_2024}. On average, each WFM pointing was observed for 15 epochs, whereas each NFM pointing was only observed twice. However, the footprints of the NFM pointings are fully covered by the central WFM pointings as well. For the WFM data, cadences between epochs ranged from less than 1~h to several months, while for the NFM data, the cadence ranged from several months to a year. All observations were performed using the nominal wavelength range, covering $4\,750-9\,350$\r{A} with a spectral resolution of $R\sim1\,800-3\,500$. A visual representation of the WFM data available for $\omega$~Cen from MUSE is shown in Fig.~\ref{fig:MUSE_Mosaic}. 
        
        As can be seen from Fig.~\ref{fig:MUSE_Mosaic}, the GTO survey targets different areas in the central region of the cluster. We note that the data discussed in this work form part of the \textsc{oMEGACat} project, recently presented in \citet{Nitschai_2023}. However, whereas the catalog published by \citeauthor{Nitschai_2023} contains the results obtained by averaging over the individual epochs, here we focus on the single-epoch data and study the stars for radial velocity variations between the individual epochs.
        
    \subsection{Reduction}
        \label{Subsec:Reduction}
        
        All observed data was processed with the standard MUSE pipeline \citep{Weilbacher_2020}, which performs all the steps necessary to reduce integral-field spectroscopic data. Following bias subtraction, signal tracing, flat fielding, and wavelength calibration, the data from each of the 24 spectrographs are stored in a pixtable. In the next step, the data from the 24 pixtables are combined, corrected for differential atmospheric refraction, and flux calibrated. As outlined in \mbox{\citet{Husser_2016}}, we do not perform a sky subtraction nor a correction for telluric absorption. These features are corrected for during the post-processing steps described below. This process is repeated for every individual exposure before the data for the three (or four, in the case of NFM data) exposures per pointing and epoch are combined and resampled onto a final data cube, showing the RA and Dec coordinates along the spatial dimensions and the wavelength along the spectral dimension. We note that we do not investigate variability between the individual exposures entering a data cube. These exposures were usually taken back-to-back, with time offsets of a few to a maximum of 30 minutes.
%        \ska[The spectral data from each of the 24 CCDs across different exposures are then combined into a single data cube showing the RA and Dec location of the source, with a complete spectrum for each source.]{}
        
\section{Analysis}
    \label{Sec:Analysis}
    
    \subsection{Spectral extraction}
        \label{Subsec:Spectral_Extraction}
        
        %The data is then matched with a photometric reference catalogue, the ACS survey of Galactic globular clusters \cite{Sarajedini_2007}, which lists the brightness and relative positions for each star in the field of view. This is to constrain the RA and Dec position for each source before the deblanding process. To deblend, the point spread function (PSF) is 'guessed' and improved with the signal to noise ratio until a final PSF is determined, at which this is applied to the data and the individual stars are resolvable and deblended. At this point, the individual spectra of each star can be extracted
        
        The extraction of individual spectra is done using the \textsc{PampelMuse} software detailed in \cite{Kamann_2013}. In order to properly perform the extraction, the MUSE data must first be matched to an astrometric reference catalog, in order to obtain relative positions and brightness estimates for the stars in the field of view. For the central fields, we used the \textit{Hubble} space telescope (HST) photometric catalog from the ACS survey of Galactic globular clusters \citep{Sarajedini_2007, Anderson_2008}. MUSE pointings that were not completely covered by the footprint of the ACS survey were instead processed with the HST catalog published by \citet{Anderson_2010}.
        
        The analysis starts with an initial guess of the point spread function (PSF), modelled as a Moffat profile, which is iteratively improved using the brightest and most isolated stars in the sample. Within each iteration, the contributions from other nearby stars are subtracted using the PSF model. The iterative process is then repeated on each layer of the datacube and each solution is combined to create a wavelength dependent PSF model and to predict the position of each source as a function of wavelength. Finally, this wavelength dependent model is used to build the final spectrum for each source by extracting the flux from each layer of the datacube.
        
        %This PSF is then fit to all stars in the sample and PSF is scaled depending on their relative brightness. The observed intensity is then subtracted from the fit and the residual is determined. The model is then improved and the process repeated until the PSF model is optimised.

        %the psf and position of the sources are modelled for each wavelength, which is then improved and optimised and then final psf model for each wavelength is applied to each model to get final value for fluxes in each wavelength - data is always in the background

        %model composed of different psf models scaled at different positions is fit to the data 
        
        %. By fitting the mock image with the observed data, the PSF model is then optimised. The spectrum for each star is then determined by modelling the PSF for each layer of the datacube, then building the spectrum by determining the intensity of the source in each layer
        
%        The mock image is cross-correlated against the observed data set and, by using a least-squares optimisation on the residual of the spectra, the PSF models are improved. When the best fitting PSF models are determined, the spectra for the individual stars can then be extracted.

%        The psf image is then optomised by fitting to the data

%        spectrum is determined by fitting the PSF to each layer of the data cube, and then building the spectrum using the intensity of the PSF in each layer 

    \subsection{Spectral analysis}
        \label{Subsec:Spectral_Analysis}
        
        %After the spectra has been extracted, the information on the stars radial velocity, age and metalicity can be determined. In order to determine the radial velocity, the measured spectra are compared to the known position of spectral emission lines. By comparing the spectra, the extent of the red/blue shifting can be determined, which can then be used to calculate the stars radial velocity. In order to determine the physical characteristics of the star, each source is compared to an isochrone of $\omega$ Cen.

        The analysis of the spectra is performed with \textsc{Spexxy} \citep{Husser_2016}, which performs full-spectrum fitting against a library of template spectra in order to measure stellar parameters and radial velocities. In order to obtain initial guesses for the former, the photometry of the stars with extracted spectra is compared to an isochrone from the PARSEC database \citep{Bressan_2012}, which assumes a distance of 5.3 kpc, extinction of $A_{V}=0.4$, an age of 13.5 Gyr and a metallicity of [Fe/H]~=~-1.2. Using a nearest-neighbour approach, each star is assigned an initial value for effective temperature (T$_{\text{eff}}$), surface gravity (log~$g$) and metallicity. We note that, while the main population of $\omega$~Cen is more metal poor ([Fe/H]~$\sim-1.7$) than our chosen isochrone, the metallicity of the latter is selected to account for the spread in metallicity across the cluster. This initial value is also improved for each star through the spectral fitting process, as discussed below. The only factor that does not change during the spectral fitting process is the log~$g$ value, which has little dependence on metallicity. Using the initial stellar parameters, a template from the G\"{o}ttingen Spectral Library of PHOENIX spectra \citep[\textsc{GLib}][]{Husser_2013} is chosen for each star and initial values for the radial velocities are determined by cross correlating the observed spectra of each star with the selected template. 

        During the actual fitting using \textsc{Spexxy}, the initial guesses for the parameters are improved using a least-squares optimisation between spectra and templates across the entire MUSE wavelength range. The details of this process vary, depending on the evolutionary state of each star. For the majority of stars (on the main sequence, subgiant, red giant, and asymptotic giant branches), the template spectra are taken again from the \textsc{GLib} library and all aforementioned parameters except for the surface gravity log~$g$ are optimized. However, the \textsc{GLib} templates do not cover the effective temperature range $>15\,000\,{\rm K}$ required to fit the extended horizontal branch stars. Therefore, these stars were instead compared to a dedicated library, recently presented in \citet{Latour_2023}. In such cases, the parameters that were optimized during the fit were T$_{\text{eff}}$, log~$g$, the helium abundance instead of the metallicity (given the absence of any significant metal lines in the MUSE spectra of hot stars) and the radial velocity.

        For each measured radial velocity, \textsc{Spexxy} returns the nominal uncertainty derived from the covariance matrix. As previous MUSE radial velocity studies have shown \mbox{\citep[e.g.,][]{Kamann_2016,Nitschai_2023}}, \textsc{Spexxy} tends to underestimate the true velocity uncertainties. As underestimated uncertainties can artificially inflate the measured binary fractions, we followed the approach outlined in \mbox{\cite{Kamann_2016}} and determined a correction factor for the uncertainty of each velocity measurement. For each measurement of each star, we selected a comparison sample of 100 measurements from the same observation with similar stellar parameters and spectral S/N. For the measurements in the comparison sample, we calculated the normalized (by the squared sum of the uncertainties) velocity differences relative to other epochs and used the standard deviation of the resulting distribution as correction factor. For more details about this process, we direct the reader to \cite{Kamann_2016}.

    \subsection{Separating CMD regions}
        \label{Subsec:seperating_CMD}

        In addition to determining the radial velocities and stellar parameters, we also identify the evolutionary stage of each star in order to investigate the binarity of each evolutionary type. To do this, we separate the regions of the CMD by eye into eight main stages: main sequence (MS), turn off (TO), sub giant branch (SGB), red giant branch (RGB), horizontal branch (HB), asymptotic giant branch (AGB) and BSS.
        
        This is an important metric, as some regions of the CMD are more likely to show binarity than others, whether it be through physical properties or observational biases. For example, BSSs are more likely to be identified as binary because this class of objects is predicted to be formed through the interactions between the two components in a binary, as discussed in Sec.~\ref{Sec:Intro}. Bright stars, such as RGB stars, tend to have a higher SNR and therefore typically have lower observed velocity errors. In our sample, RGB stars have a mean velocity error of $3$~kms$^{-1}$ compared to MS stars, which have a mean velocity error of $12$~kms$^{-1}$ and lower velocity errors improve the detection limits of radial velocity variations.

        We note that due to the nature of the assignment of evolutionary stage, there are some cases where multiple stages are applicable, namely at the edges of connecting regions. For these cases we include this source for all applicable evolutionary stages. This may mean that the number of stars and binary systems reported totaled over each evolutionary stage may be larger than the reported global values. 
        
    \subsection{MUSE sample selection criteria}
        \label{Subsec:Final_Sample}

        We perform a number of quality cuts on the MUSE sample as well as a cluster membership selection. The details of this process are provided in the following:

        \begin{itemize}
            \item Only results derived from spectra that were formally successfully fitted by \textsc{Spexxy} were considered. A common reason for the failure of a spectral fit was that it hit the limits of the template grid for T$_{\text{eff}}$, log~$g$, or [Fe/H].
            \item The contamination from nearby sources -- that were too close to the target star to be resolved in the MUSE data and hence could not be deblended by \textsc{PampelMuse} -- in the extracted spectrum is less than 5\%.
            \item The star was extracted more than 2 spaxel away from the edge of the MUSE field of view.
            \item The magnitude accuracy of the spectrum, as determined by \textsc{PampelMuse}, is above 0.6. To determine this value, \textsc{PampelMuse} tries to recover the magnitudes available in the photometric reference catalog by integrating over the extracted spectra. The magnitude accuracy is a measure for the agreement between the two, relative to the scatter observed for spectra of similarly bright stars. A value of 1 indicates perfect agreement whereas a value of 0 indicates a strong outlier.
            \item The reliability of each radial velocity measurement, as defined by \citet{Giesers_2019}, is over 80\% 
            \item The values derived from the spectrum analysis for T$_{\text{eff}}$, log~$g$, and [Fe/H] show no outliers when compared to results from other spectra obtained for the same star.
            \item Field stars were discarded using a membership probability cut of 0.5. As explained in \citet{Kamann_2016}, the probability of cluster membership is derived by comparing the average [Fe/H] and radial velocity measurements of each star to assumed populations of cluster and Milky Way stars.
        \end{itemize}
                
\begin{figure}
            \centering
            \includegraphics[scale=0.48]{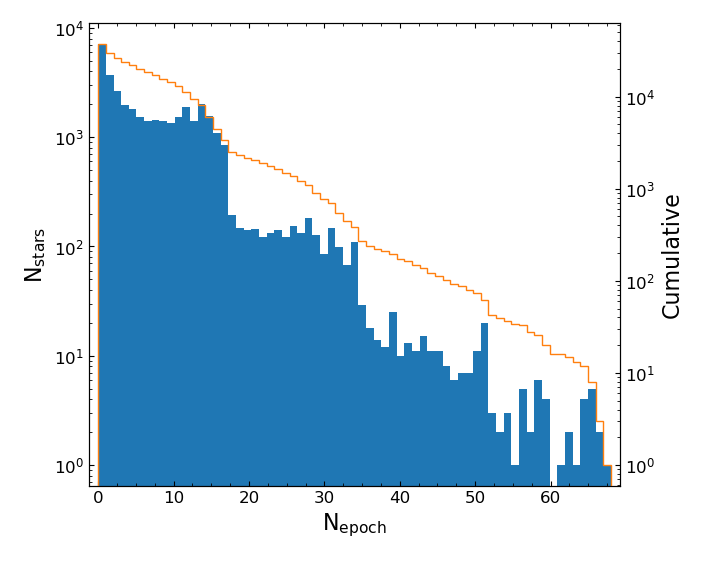}
            \caption{The distribution of the number of epochs for each source in $\omega$~Cen. The median number of epochs is 6. The orange histogram represents the cumulative number of stars that have at most a given number of epochs.}
            \label{fig:epoch_frequency}
        \end{figure}

        Following these cuts, there are 312\,248 individual spectra for a total of 37\,225 stars. The distribution of the number of epochs across the sample is shown in Fig.~\ref{fig:epoch_frequency}. On average, a star in our sample has valid radial velocity measurements for 6 epochs. However, a large scatter in the available number of epochs per star is evident in Fig.~\ref{fig:epoch_frequency}. This is caused by the varying observing conditions during the campaign, impacting the number of spectra that could be extracted from the MUSE data during each epoch. The sharp cutoff beyond 17 epochs indicates the maximum number of epochs available per pointing. Note, however, the tail extending to $\gtrsim 50$ epochs, composed of stars that are located in the overlap regions between adjacent pointings.

\section{Variability Measurement}
    \label{Sec:Variability_Meaurement}

    \subsection{Identifying photometrically variable stars}
        \label{Subsec:Variable_Stars}

        In order to properly identify binary candidates in $\omega$ Cen, we also must remove stars that have been previously reported as photometrically variable, such as RR Lyrae and SX Phoenicis (SX Phe) stars. These radially pulsating stars show radial velocity variations, however the variation is not caused by a companion orbiting around in a binary. In this study we compare our sample to a compilation of two catalogues of photometric variables available in the literature, a catalogue from \cite{Clements_2001} and a catalogue from \cite{Braga_2020}.

        The \citeauthor{Clements_2001} catalogue is a summary of literature reporting variable stars, such as RR Lyrae, SX Phe stars, Semi Regulars (SRs) and Long Period Variables (LPVs). The sources in the \citeauthor{Clements_2001} catalogue were matched to the MUSE data initially via their WCS positions, with a 2 arcsecond allowance. To avoid mismatches, we restricted the potential matches in the MUSE catalog to stars on the Horizontal Branch (HB), Blue Straggler Stars (BSS), and Asymptotic Giant Branch (AGB) stars, as the types of variable stars previously mentioned are only found in these areas of the colour-magnitude diagram (CMD). The matches were then manually confirmed by comparing the magnitudes from the \citeauthor{Clements_2001} catalogue to the F625W magnitudes available in the \citet{Anderson_2010} photometry. A source was considered a match if the magnitude reported in the \citeauthor{Clements_2001} catalogue was within 1 magnitude of the HST magnitude, and we confirmed that all matched sources have a 2D separation within 0.5$\arcsec$ between the reported RA and Dec from the MUSE and \citeauthor{Clements_2001} catalogues. Because of the good astrometric and photometric agreement and the pre-selection based on specific regions of the CMD, we are confident that all sources from the \citeauthor{Clements_2001} and \citeauthor{Braga_2020} catalogues are correctly matched to their MUSE counterpart. From the \citeauthor{Clements_2001}, we identify a total of 18 variable stars with a MUSE counterpart in our sample. The \citeauthor{Braga_2020} catalogue contains photometric variables, such as RR Lyrae, Type II Cepheids and LPVs that were found using Near Infrared and Optical data. The \citeauthor{Braga_2020} catalogue was also matched to the MUSE data through coordinate matching. From this catalogue, we identified a further 4 photometric variables.

%        \textcolor{magenta}{delta ra, delta dec and delta magnitude plots, add average ra dec and mag offsets in paper}

%        In the \citeauthor{Clements_2001} catalogue, there are a total of 32 variables in the MUSE field of view that have a reported variable counterpart. This includes 21 RR Lyrae stars, 4 SX Phoenicis stars, 2 Semi Regular variables (SRs), 2 Long Period Variables (LPVs), and 2 undefined variables.
%        \ska[and so could not be confirmed with any MUSE source]{}.
        
        The locations in the CMD of the matches to both the \citeauthor{Braga_2020} and \citeauthor{Clements_2001} catalogues are visualized in the right panel of Fig.~\ref{fig:CMD}. As expected, LPV and SR are found on the AGB branch, the RR Lyrae stars cluster on the horizontal branch, and nicely illustrate the location of the instability strip, and SX Phe stars are found among the blue stragglers, in agreement with the findings by \citet{Cohen_2012}.
        
        \subsubsection{MUSE defined photometric variables}
            \label{Subsubsec:MUSE_photo_variables}
    
    %        \ska[The MUSE data contained a separate measurement of photo variability determined by constructing magnitudes for each observation and measuring the variation \mbox{\citep{Kamann_2018}}.]
            The MUSE data offer the possibility to search for photometric variables, by reconstructing broadband magnitudes from the extracted stellar spectra and comparing them across different epochs. To account for variations in the observing conditions, sets of comparison stars are defined within each data cube that is processed. Using this method, which is detailed in Appendix~A.1 of \citet{Giesers_2020}, we assigned each star a probability to show photometric variations. For the purposes of this study, we select a cut of 0.8 for the photometric variability, where stars with a photometric variability $>0.8$ are considered photometrically variable. The stars that we class as photometrically variable are shown as black points on the left panel of Fig.~\ref{fig:CMD}. In total, 912 stars were listed as photometric variables, 6 of which match the variable stars reported in the literature. As can be verified from Fig.~\ref{fig:CMD}, the majority of RR~Lyrae stars are identified as photometrically variable in the MUSE data, which can be explained by their relatively strong brightness variations compared to other types of variables. However, we emphasize that the majority of stars identified as photometrically variable in the MUSE data are unlikely to show intrinsic variability. Instead, it is more likely that the photometric variations point to problems during the extraction of the spectra. Therefore, we exclude the stars showing high probabilities for photometric variability from further analyses.

    \subsection{Calculating the probability of variability}
        \label{Subsec:Probability_calc}
        
        A common way to detect velocity variations is to determine the $\chi^2$ of the velocity measurements per star under the assumption of a constant actual velocity. However, when processing samples as large as ours, a considerable number of large $\chi^2$ values is expected even in the absence of binaries. This statistical noise presents a challenge when determining which stars are variable, and can strongly impact the results, especially in the regime of low binary fractions expected for $\omega$~Cen. We utilise the probability calculation method presented in \cite{Giesers_2019}, as this method avoids the high false positive detection rate by weighting each $\chi^2$ value against the likelihood of the source being statistical noise. In practice, the number of stars detected at or above a given $\chi^2$ value is compared to the number expected from statistical noise, and their fraction is converted to a probability value, henceforth $P_{\rm var}$. To calculate uncertainty, we use a bootstrapping analysis with a $1\sigma$ confidence interval.

        % \textcolor{magenta}{Using the $4\sigma$ limit from the study by \citet{Sana_2013}, we find approximately 835 stars with radial velocity variations. Using an unweighted chi square test, however, increases the instance of false positive detections}.
        
\begin{figure*}
            \centering
            \includegraphics[scale=0.53]{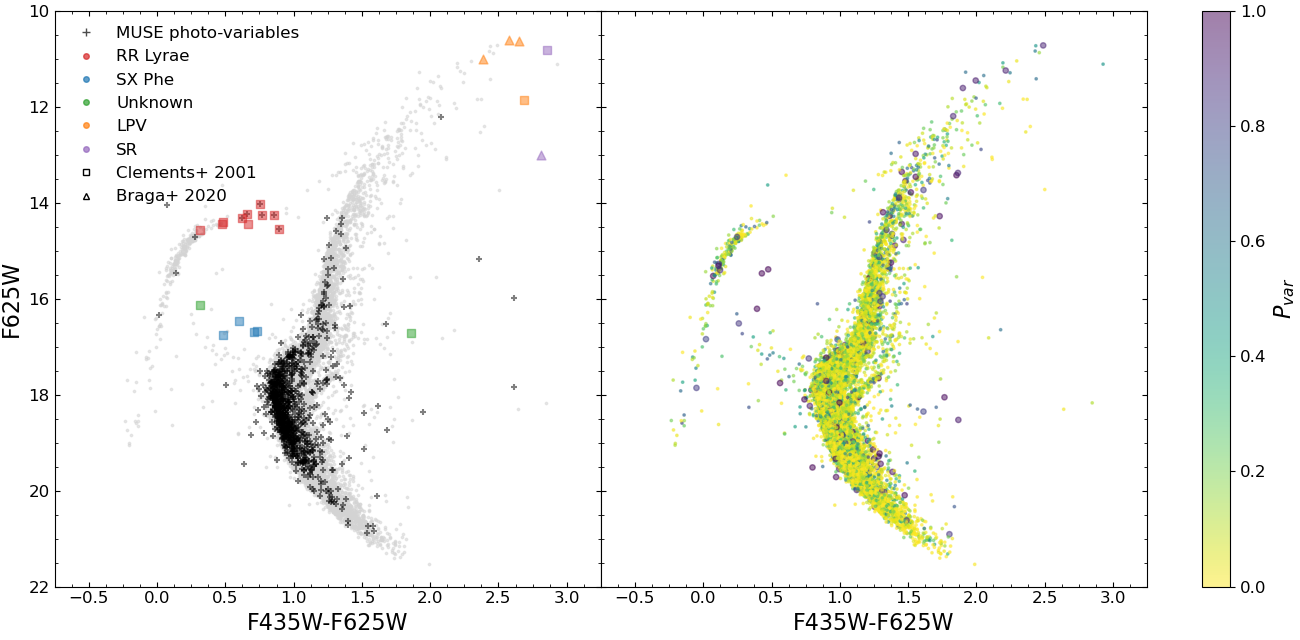}
            \caption{Stellar variability in $\omega$ Cen. Both panels show the distribution of stars in the MUSE sample across the colour magnitude diagram (CMD) of the cluster using F625W and F435W magnitudes. The left hand panel shows stars that are considered photometrically variable either in the MUSE data or the \citet{Clements_2001} and \citet{Braga_2020} catalogs. The right panel shows the sample with all photometic variables removed and uses a colour overlay to indicate $P_{\rm var}$, the probability of a star to show velocity variations, where purple is more likely to be variable and yellow less likely, as indicated by the colour bar to the right. The larger points of the right hand panel show stars with $P_{\rm var}>0.8$, a cut which is discussed in Sec.~\ref{Subsec:Cut_value_Dependence}, which we consider as binary stars.}
            \label{fig:CMD}
        \end{figure*}

        We show the calculated $P_{\rm var}$ values for each star in the right panel of Fig.~\ref{fig:CMD} across the different regions of the CMD. We find a particularly high concentration of potential binary stars among blue stragglers, however this will be discussed throughout Sec.~\ref{Sec:Results}.

\begin{figure}
            \centering
            \includegraphics[scale=0.5]{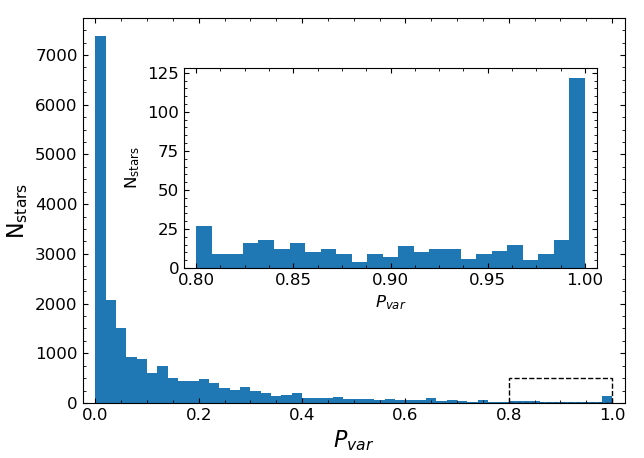}
            \caption{The number of stars in the MUSE sample with six or more epochs as a function of variability probability, $P_{\rm var}$, as designed by \citet{Giesers_2019}. The inset axis shows the distribution of $P_{\rm var}$ between 0.8 and 1.}
            \label{fig:Probability_Graph}
        \end{figure}

        The distribution of $P_{\rm var}$ is also shown in Fig.~\ref{fig:Probability_Graph}. Note that only stars with a minimum of six measurements are included in Fig.~\ref{fig:Probability_Graph}, a choice that will be motivated in Sec.~\ref{Subsec:Cut_value_Dependence} below. The inserted axis shows the distribution of stars with a $P_{\rm var} >0.8$, which will also be discussed in Sec.~\ref{Subsec:Cut_value_Dependence}. It is evident from Fig.~\ref{fig:Probability_Graph} that the $P_{\rm var}$ distribution returned by the method of \citet{Giesers_2019} is bimodal, with a dominant peak at low probabilities. We will further elaborate this point in the following sections of this paper. We also present the full binary fraction in Sec.~\ref{Subsec:Results_Disc_frac}, after a discussion on the chosen cut values.

\section{Completeness Correction}
    \label{Sec:Completeness_Correction}

    In order to determine the binary fraction of $\omega$~Cen, we need to correct for the incompleteness in our analysis. Inevitably, there will be binary stars that are not classified as variable via the probability method detailed in Sec.~\ref{Sec:Analysis} due to effects such as systems having a low inclination, low-mass companions on wide orbits, or damping of the true radial velocity curves by secondary stars of a similar luminosity (cf. Sec.~\ref{Subsec:Simulation_Setup}). 

    Although evolutionary models for globular clusters exist, such as MOCCA \citep[MOnte Carlo Cluster simulAtor;][]{Giersz_1998} and CMC \citep[Cluster Monte Carlo;][]{Joshi_2000}, there are no evolutionary models of $\omega$~Cen with a realistic binary population to date. There are two main factors that make modelling $\omega$~Cen difficult, the first being the unknown origin of $\omega$~Cen which, as a theorised former NSC, would require models that rely on too many assumptions to realistically track evolution. The second issue is the high number of stars in $\omega$~Cen, which would make realistic models too computationally expensive to run.

    Therefore, to correct for these effects we generate mock MUSE samples, with known binary populations and individually known orbital parameters, and apply the same probability calculation method as was applied to the observed data. By determining the ratio of the inputted binary fraction to the outputted discovery fraction of the simulated cluster, a correction factor can be derived and applied to the actual data in order to obtain a completeness-corrected binary fraction. 

    To generate the mock samples, we use the observed time stamps and velocity errors for each measurement of each star and calculate a new set of velocity measurements for each star, with variations depending on whether the star has been simulated to have a binary companion or it is considered a single star.

    \subsection{Physical principles}
        \label{Subsec:Physics_Principles}

        In order to create mock samples that are representative of $\omega$ Cen, there needs to be some consideration of what types of binary stars are expected in the dense cluster environment. In particular, we consider the constraints imposed by three-body interactions and Roche-lobe overflow on the distribution of orbital periods we expect to observe.

        \subsubsection{Binary hardness}
            \label{Subsubsec:Binary_Hardness}
        
            The first point to consider is the hardness of the binary system; the binding energy between a primary star and its secondary companion relative to the kinetic energy of an average cluster member. If a binary system has a low binding energy (a soft binary), it will be easily disrupted  by interactions with single stars or other binary systems. If a binary system has a high binding energy (a hard binary), it is much more likely to survive the cluster environment, as it is harder for interactions to destroy the system \citep{Heggie_1975}. In order to determine the hardness of a binary, we follow \citet{Ivanova_2005} and write down the ratio between the binding energy of a binary and the kinetic energy of a typical clusters star as
        
            \begin{equation}
                    \eta = \frac{GM_1M_2}{a\sigma^2\langle M \rangle}\,, %> 1 \text{ for hard binaries}
            \label{eqn:Binary_Hardness}
            \end{equation}

            where $\eta$ is the binary hardness, $M_1$ and $M_2$ are the masses of the primary and secondary stars respectively, $\langle M \rangle$ is the average mass of a star in the cluster, $a$ is the semi-major axis of the binary system, and $\sigma$ is the average velocity dispersion of the cluster.
            
            A binary system is considered a soft binary if $\eta<1$, and a hard binary if $\eta>1$. Using Eq.~\ref{eqn:Binary_Hardness} with Kepler's third law, the binary hardness can be determined for different orbital periods for given companion masses. For example, assuming a dispersion of $\sigma=21\,{\rm km\,^{-1}}$, calculated using the MUSE data, and an average mass of $\langle M \rangle=0.5\,{\rm M_\odot}$, we can estimate that the maximum period for a binary system composed of two typical main sequence stars of average mass is approximately $10^3$ days, with the maximum period getting smaller as the companion becomes less massive. As soft binary systems are expected to be rapidly disrupted in the dense environment of $\omega$ Cen, simulated binary systems with a hardness $\eta\leq1$ are ignored.
            
%            The binary hardness as a function of orbital period for different companion masses of a typical 0.8M$_\odot$ main sequence star is shown in Figure \ref{fig:Binary_Hardness}.
            
%            \begin{figure}
%                \centering
%                \includegraphics[scale=0.58]{Images/Binary_Hardness.png}
%                \caption{The distribution of binary hardness with different companion masses. Each trend shows the hardness of a system with a primary mass of 0.8M$_\odot$ and a varying companion mass as indicated. The dashed line at $y=1$ represents the limit for a system to be classified as a soft binary system, where hardness$<1$.}
 %               \label{fig:Binary_Hardness}
 %           \end{figure}

        \subsubsection{Roche lobe filling}
            \label{Subsubsec:Roche_Lobe_Filling}
        
            In order to obtain a lower limit for the periods feasible in our sample, we consider the Roche limit, i.e. the point at which two stars in a binary system become connected by the mass transfer of one of the stars. The radius $r_1$ of the Roche lobe of the primary star is described by the \cite{Eggleton_1983} equation:
            \begin{equation}
                \frac{r_1}{A} = \frac{0.49q^{2/3}}{0.6q^{2/3}+\text{ln}(1+q^{1/3})}, \quad with \quad q = \frac{M_1}{M_2}
            \end{equation}
            \label{eqn:Eggleton}
            Here, $A$ is the orbital separation of the stars and $M_1$ and $M_2$ are the mass of the primary and secondary star, respectively.
            
            In order to determine if a binary system is feasible, we compare $r_1$ to the radius of the primary star, estimated using the isochrone comparison described in Sec.~\ref{Subsec:Spectral_Analysis} above. If the latter exceeds the former, we discard the binary system and treat the system as a single star. We note that interacting binary stars showing Roche-lobe overflow are expected in a dense environment such as $\omega$~Cen, but are unlikely to contribute significantly to the overall statistics. The study by \citet{Goettgens_2019} found only 9 emission-line sources in the MUSE sample of $\omega$~Cen, with most of them being known photometric variables (cf. Sec.~\ref{Subsec:Variable_Stars}).

%            \begin{equation}
%                R = \left(\frac{L}{4\pi\sigma_{sb}T^4}\right)^{\frac{1}{2}} \quad with \quad L = L_{\odot}10^{L_r}
%            \end{equation}
%            \label{eqn:radius}

%            \ska[Where $R$, $L$ and $T$ are the radius, luminosity and effective temperature of the primary star taken from the isochrone. If the stars radius is greater to then or equal to the radius of the Roche lobe, this therefore means that the system is considered to be a single star.]{}
%             L = constants.L_sun*(10**(L_r))
 %   return((L/(4*math.pi*constants.sigma_sb*(T**4)))**(1/2)).decompose().to(u.km)

  %  (L/4pisigmaT4) 1/2
            
    \subsection{Generating mock samples}
        \label{Subsec:Simulation_Setup}

        When creating a mock sample, we pick a random subsample of the stars (with a size corresponding to the selected binary fraction) and assign each star as the primary star in a binary systems with orbital properties in line with the following assumptions:

        \begin{itemize}
            \item The mass ratio distribution is described by a power law function $F(q) = q^{\gamma}$. The default value for $\gamma$ is 0, corresponding to a uniform distribution of mass ratios, in line with \citet{Ivanova_2005} and \citet{Wotias_2001}.
            \item The default distribution for inclination is a uniform distribution of $\arccos{i}$ between 0 and 1, corresponding to an isotropic distribution of orbital planes.
            \item The period distribution is a log normal distribution with a mean value of $10^{0.5}$ days and a standard deviation of $10^{1.5}$ days. This choice is motivated by the period distributions found in the simulations of \citet{Ivanova_2005}.
            \item The phase of time in the orbit during the first epoch, t$_0$, is a uniform distribution between 0 and the value of P for each given binary.
            \item The default distribution of eccentricity, e, is a beta distribution, with default values of $\alpha=2$ and $\beta=5$. A beta distribution was used motivated by work by \citet{Kipping_2013} and we test a wide range of assumed values (discussed in Sec.~\ref{Subsec:Dependence_on_Orbital_Parameters}). For short periods, we also implement a maximum eccentricity to avoid unphysical orbits, according to the following formula \citep{Moe_2017}.
            \begin{equation}
            \text{e}_{\text{max}}(\text{P})=1-\left(\frac{\text{P}}{\text{2 days}}\right)^{-2/3} \quad \text{for P > 2 days}
            \label{eqn:e_max}
            \end{equation}
            If a binary system is found to have an eccentricity higher than e$_{\text{max}}$, that systems eccentricity is set to e$_{\text{max}}$. For periods $\leq 2$ days, we assume that the orbit is circularised.
            \item The argument of periapsis, $\omega$, is a uniform distribution between 0 and $2\pi$ for each given orbit.
        \end{itemize}

        Along with each set of randomised orbital parameters for the binary systems, each star, whether binary or not, is assigned a randomised mean system velocity, v$_{\text{sys}}$, which is taken from a normal distribution with the mean value $\mu=250$kms$^{-1}$, the mean velocity of $\omega$ Cen calculated using the MUSE data, and a standard deviation $\sigma=21$kms$^{-1}$, the central velocity dispersion of $\omega$ Cen, also calculated using the MUSE data. We assume a constant velocity dispersion as the velocity dispersion curve of $\omega$~Cen is relatively shallow \citep[e.g.,][]{Solima_2019} within the central region relevant for this study, and has a limited effect on the results.
        
        Following the assignment of orbital parameters, we discard all systems that violate the criteria on binary hardness and Roche-lobe overflow set out in Secs.~\ref{Subsubsec:Binary_Hardness} and \ref{Subsubsec:Roche_Lobe_Filling}, respectively. Theses stars are treated as single stars instead. We note that the binary fraction that we assume for the mock set is calculated following this step. For each actual velocity measurement in the MUSE catalog, we then create a mock velocity measurement. To this aim, we first use the orbital parameters of the mock sample to determine the Keplerian velocities of the stars in binaries, at the timestamps of the observations.

% 
%             \begin{equation}
%                 \text{e}_{\text{max}}(\text{P})=1-\left(\frac{\text{P}}{\text{2 days}}\right)^{-2/3} \quad \text{for P > 2 days}
%                 \label{eqn:e_max}
%             \end{equation}

%            \ska[Where e$_{\text{max}}$ is the upper limit of eccentricity, and P is the orbital period in days. If a binary system is found to have an eccentricity higher than e$_{\text{max}}$, that systems eccentricity is set to e$_{\text{max}}$. For periods $\leq 2$ days, we assume that the orbit is circularised.]{}

%            \ska[After creating the randomised parameter set for each star, the simulation then removes unphysical systems from the binary population by performing a series of checks. The checks are based on the binary hardness and Roche lobe filling described in Sections \ref{Subsubsec:Binary_Hardness} and \ref{Subsubsec:Roche_Lobe_Filling} respectively. If either test fails (the binary hardness is less than 1 or the Roche lobe radius is smaller than the stellar radius), then the star is treated as a single star.]{} 

            Having calculated the Keplerian velocity, we then consider luminosity damping of a primary star due to its secondary, particularly in the case where the primary and secondary stars have comparable luminosities. In these cases, the red and blueshifts in the spectra of the two components relative to the observer negate one another, resulting in widened spectral lines rather than observable shifts. As shown by \citet{Bodensteiner_2021} or \citet{Saracino_2023}, this effect causes the detection efficiency of MUSE observations to drop for binaries composed of unevolved stars with mass ratios close to unity. To take this effect into account, we adopt the equation used by \cite{Giesers_2019}
            \begin{equation}
                v_{\text{obs}} = v_{\text{sys}} + (1 - 10^{0.4(m_1-m_2)})v_{\text{t}}\,, 
            \end{equation}
            \label{eqn:Damping_Factor}
            where $v_{\text{obs}}$ is the observed velocity of the primary, $v_{\text{sys}}$ is the systemic velocity of the binary, $v_{\text{t}}$ is the true Keplerian velocity of the primary, and $m_1$ and $m_2$ are the F625W magnitude of the primary and secondary star, respectively. The magnitude of the secondary is adopted from the isochrone, by finding the closest match in terms of mass along the main sequence. This damping factor is then applied to the Keplerian velocities of all binary systems in the mock data.
            
            For the single stars in the sample, the velocity is simply set to $v_{\rm sys}$ for all timestamps, which is a valid assumption as the detection efficiency is not impacted by the exact velocity measurement. Finally, we create each mock measurement by drawing a random number from a Gaussian distribution, centred on the velocity calculated for the star and the timestamp in question, with standard deviation equal to the uncertainty of the observed velocity. These mock data sets are created for a set of 11 input binary fractions ranging between 0\% and 20\% input fraction in steps of 2\%.

%            \ska[If a star passes the physicality tests, it then goes through orbital velocity calculations. For each star, each observation date and observed velocity error is taken. With these values as a framework and using the randomised parameters, a new velocity amplitude is calculated for each source. Using the velocity amplitude, the randomised orbital parameters, and the randomised systemic velocity of the binary system $v_{\text{sys}}$, value is calculated for the radial velocity of the mock system.]{} 

\begin{figure}
                    \centering
                    \includegraphics[scale=0.5]{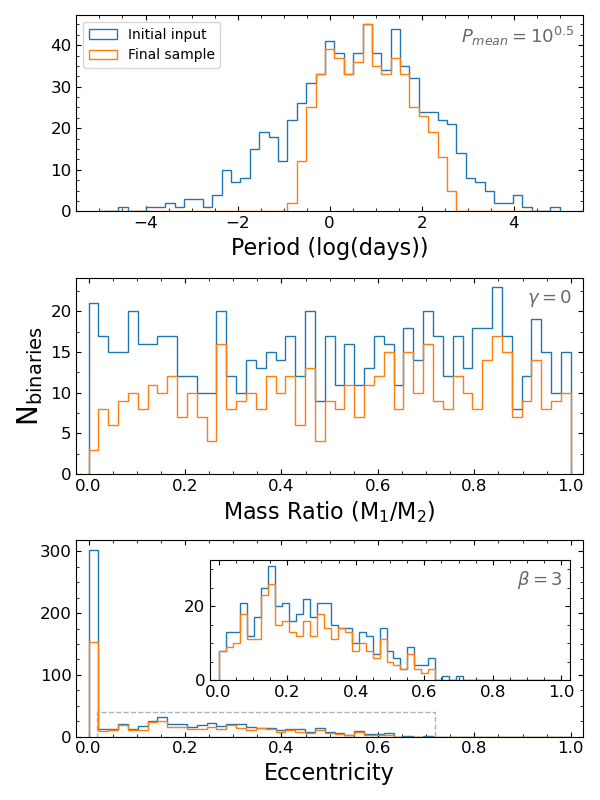}
                    \caption{The distribution of orbital parameters in the mock sample with an intrinsic binary fraction of 4\%, both before (blue) and after (orange) the removal of soft and interacting binaries. The top panel shows the distribution of periods, the central panel the distribution of mass ratios, and the bottom panel the distribution of eccentricities. In order to better visualize the distribution of binaries with non-zero eccentricity, the inserted panel shows the eccentricity distribution following the removal of round orbits. The initial sample (blue) contains 485 stars and the final sample (orange) contains 265 stars.}
                    \label{fig:Parameter_distribution}
                \end{figure}

            Figure~\ref{fig:Parameter_distribution} shows the distribution of the orbital parameters in the baseline case, for which $\log(P_{\text{mean}}/\text{d})$=0.5, $\log(P_{\sigma}/\text{d})$=1.5, $\gamma$=0, $\beta$=3, and with input binary fraction $4\%$, for both the initial input of binary systems, and the surviving fraction following the removal of soft and interacting binaries. We select the input fraction as $4\%$ as this gives a discovery fraction comparable to the observed data. The top panel of Fig.~\ref{fig:Parameter_distribution} shows that the Roche-lobe criterion outlined in Sec.~\ref{Subsubsec:Roche_Lobe_Filling} limits the periods we expect in our sample to $\gtrsim0.1~\text{d}$, whereas the criterion on binary hardness defined in Sec.~\ref{Subsubsec:Binary_Hardness} confines the expected distribution to $\lesssim1\,000~\text{d}$. The impact of the binary hardness criterion is also visible in the middle panel of Fig.~\ref{fig:Parameter_distribution}, as binaries with low mass ratios are preferentially removed from the input sample. The peak of eccentricity at 0, visible in the lower panel of Fig.~\ref{fig:Parameter_distribution}, is due to the assumed circularization of binaries with periods under 2 days.
            %\ska[, as this is the orbital period at which orbits are assumed to have circularised \mbox{\citep{Giesers_2019}}]{}. 
                
\section{Results and Discussion}
    \label{Sec:Results}

    % \subsection{Dependence on Cut Values}
    \subsection{Binary recovery as a function of $P_{\rm var}$}
        \label{Subsec:Cut_value_Dependence}

\begin{figure}
            \centering
            \includegraphics[scale=0.54]{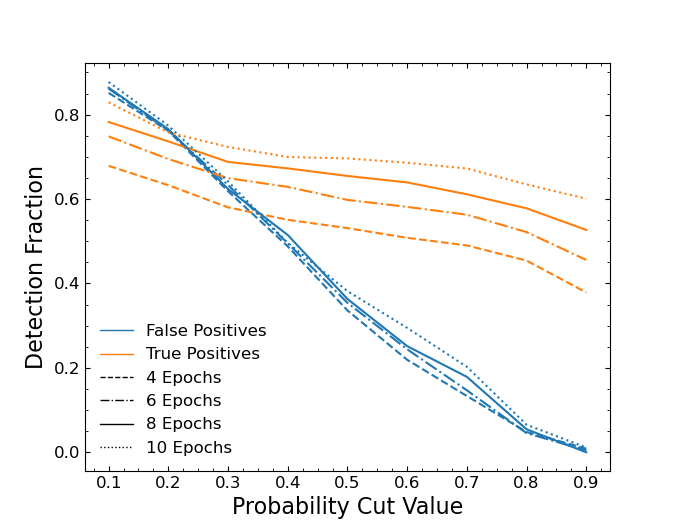}
            \caption{Completeness and purity of the sample of detected binaries in $\omega$~Cen, as predicted by the mock data. In orange, we illustrate how the fraction of detected binary stars depends on the minimum required $P_{\rm var}$, while in blue, we show the dependence of the fraction of false positives as a function of the $P_{\rm var}$ cut-off. Different line symbols are used to illustrate the impact of increasing the number of required epochs available for each star entering the analysis.}
            \label{fig:False_positive_heatmap}
        \end{figure}

        Using the mock samples, we first investigate how the completeness and purity of our sample of detected binary candidates in $\omega$~Cen changes depending on the requirements we impose on the minimum values of $P_{\rm var}$ and the number of epochs. Our aim here is to minimise the number of false positive detections of binary stars, whilst at the same time maximizing the fraction of detected binaries. In Fig.~\ref{fig:False_positive_heatmap}, we show the fraction of detected binaries (relative to the number of binaries in the mock sample) as well as the fraction of false positives (relative to the number of total detections) as a function of the cut-off on the $P_{\rm var}$, and for different cut-offs on the number of available epochs.

        Fig.~\ref{fig:False_positive_heatmap} shows that the fraction of false positives is most dependent on the minimum $P_{\rm var}$ value, while the minimum number of epochs only marginally affects the fraction. However, the minimum number of epochs has a significant impact on the completeness of our analysis, which increases by $\sim10\%$ when going from 4 to 10 epochs.\footnote{Note that ``completeness'' is defined relative to the number of stars surviving our quality and epoch cuts, not relative to the entire MUSE sample.}
        
        In order to maintain a clean sample of binary stars, we request that the false positives represent $2\sigma$ outliers in the final sample, i.e. that $\lesssim5\%$ of the stars considered as binaries are false positives. For the baseline simulation discussed at the end of Sec.~\ref{Subsec:Simulation_Setup}, adopting a minimum of $P_{var}=0.8$ yields a false positive fraction of $4.5\%$, almost independent of the number of epochs required. We note that other methods can be used to infer a well-defined cut-off value. For example, \citet{Johnson_2015} studied the relation between completeness and contamination and adopted the probability threshold that minimized the distance of the relation to the point representing the ideal situation of a fully complete and uncontaminated sample.
        
        Furthermore, we adopt a minimum number of 6 epochs per star, as it represents a good trade-off between maintaining a large sample and retaining a high fraction of correctly identified binaries. Using these cuts, we find that the fraction of known binary stars from the mock simulation that are correctly identified is $52\%$. We completed a similar test, using these updated cut values, to assess the effect of the minimum required signal-to-noise ratio (SNR) on the false positive fraction, but the effect was negligible compared to the effects of $P_{\rm var}$ and epoch cut. Having applied the 6 epoch cut to the observed data, our sample now includes 19\,059 individual stars.

    % \subsection{Dependence on Orbital Parameters}
    \subsection{Binary recovery as a function of orbital parameters}
        \label{Subsec:Dependence_on_Orbital_Parameters}

\begin{figure}
                \centering
                \includegraphics[scale=0.55]{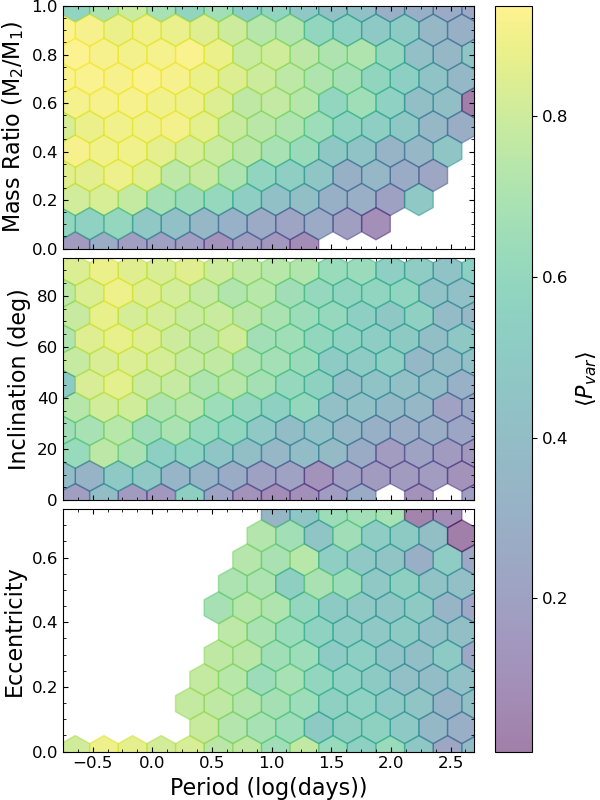}
                \caption{The mean binary probability, $P_{\rm var}$, of binary systems in the mock samples, shown as a function of their orbital period, mass ratio, inclination and eccentricity. Brighter colours correspond to higher mean values, as indicated by the colour bar to the right.}
                \label{fig:parameter_scatter}
            \end{figure}

        Using the simulated data, we can investigate the completeness of our binary search in $\omega$~Cen for different parts of the parameter space. To this aim, we show in Fig.~\ref{fig:parameter_scatter} how the detection of velocity variation is affected by the period, mass ratio, inclination, and eccentricity of the mock binary systems. We combined the values of $P_{\rm var}$ of the input binary population across all simulations with the default parameter distributions, binned them as a function of the various orbital parameters, and calculated the mean value of $P_{\rm var}$ in each bin. As expected, the period of a binary system has the strongest impact on its detectability. Excluding systems with very low mass ratios, our detection rate is expected to be nearly 100\% for periods $\lesssim10~{\rm d}$, whereas it drops to $\sim30\%$ for longer-period binaries. In terms of mass ratio, we see a significant increase in the detection rate for more massive companions. The MUSE data appear most efficient in detecting binaries with mass ratios $\sim0.6-0.8$, with the drop in efficiency as the mass ratios approach unity due the damping explained in Sec.~\ref{Subsec:Simulation_Setup}. The drop in efficiency with decreasing inclination visible in the central panel of Fig.~\ref{fig:parameter_scatter} is also expected. It is reassuring though that even for inclinations as low of 20~degrees, we still detect a considerable fraction of the binaries.
            
        In light of the strong variation of $P_{\rm var}$ across the binary parameter space expected in $\omega$~Cen that is visible in Fig.~\ref{fig:parameter_scatter}, we also investigate the impact of changing the binary parameter distributions that are underlying the mock sets. To test the impact of the assumed period distribution, we use a range of $\log(P_{\text{mean}}/\text{d})$ values between -0.6 and 2.0, in steps of 0.2, with a fixed standard deviation of $10^{1.5}$ days. The reasoning behind this choice is that the resulting period distributions include some that are significantly skewed towards the minimum period range set by the onset of Roche-lobe overflow (cf. Sec.~\ref{Subsubsec:Roche_Lobe_Filling}), and others that are significantly skewed towards the boundary between hard and soft binaries (cf. Sec.~\ref{Subsubsec:Binary_Hardness}). We also test the assumption of a log-normal distribution by generating a sample with a log-uniform distribution between $10^{-2}-10^{4}$~days, which approximates the distributions found by \citet{Ivanova_2005} for low-density clusters. For the full sample, we find that the fraction of detected binary systems decreases by $\sim0.05$, which is within the uncertainty of the global completeness reported in column (e) of Tab.~\ref{table:Binary_Fraction}. This is likely a result of the final sample of binary systems being heavily dependent on the removal of stars filling their Roche lobe and soft binaries, which will be discussed below.
        
        To test the impact of the mass ratio distribution, we use a range of $\gamma$ values (see Sec.~\ref{Subsec:Simulation_Setup}) ranging from 0.6 to 2.0 in steps of 0.2. This distribution is chosen as it allows testing of both a uniform mass distribution, which is used as an assumption in the simulations by \cite{Ivanova_2005} in line with \cite{Wotias_2001}, but also distributions biased towards higher and lower mass ratio systems respectively. To test the impact of the assumed eccentricity distribution, we use a range of $\beta$ values (described in Sec.~\ref{Subsec:Simulation_Setup}) between 3 and 11 in steps of 2, resulting in distributions which are more biased to circularised and elliptical orbits, respectively. For each parameter distribution, we determine the discovery fraction for each of the 10 input binary fractions, ranging from 0\% to 20\%.

\begin{figure}
            \centering
            \includegraphics[scale=0.565]{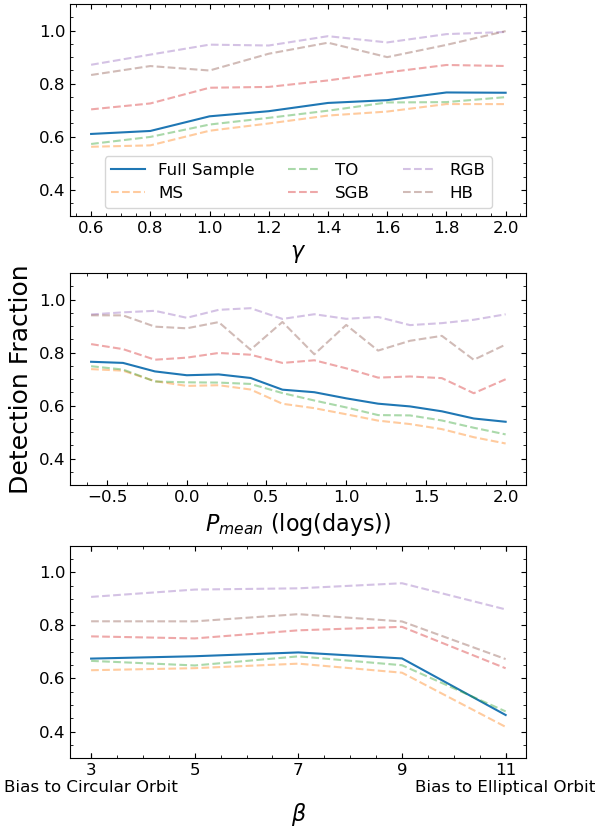}
            \caption{The recovered fraction of binaries in the mock data, as a function of the assumed values of different parameters that govern the distribution of orbital properties in the mock data: the power-law coefficient of the mass-ratio distribution (top), the mean period (middle), and the $\beta$-parameter defining the shape of the eccentricity distribution (bottom). The results for the full MUSE sample are shown as solid blue lines, while dashed coloured lines display the results obtained for different evolutionary stages.}
            \label{fig:Correction_value_plot}
        \end{figure}

        Following these tests, and having determined appropriate cut values based on $P_{\rm var}$ and the minimum number of measurements, we can now determine the completeness of the sample. Our completeness is based on the mock data with the default parameter assumptions, as detailed in Sec.~\ref{Subsec:Simulation_Setup}. To calculate the completeness, we compare the output binary fraction of binary systems detected according to Sec.~\ref{Subsec:Cut_value_Dependence} against the input binary fraction across all of the 11 input binary fractions (see Sec.~\ref{Subsec:Simulation_Setup}), and determine the gradient of the relation. To determine the uncertainty, we combine the standard deviation of the combined correction factors for all of the mock data across each of the various parameter distributions. In Fig.~\ref{fig:Correction_value_plot}, we show the dependence of the detection on the values assumed for the different distribution parameters, averaged over all input binary fractions. The detection fraction is shown for both the entire MUSE sample, and different regions in the CMD of $\omega$~Cen. From the figure, we see that the orbital period distribution has the largest effect on the completeness correction value, the latter varying by $\sim20\%$ between the two extremes of the distribution parameter range. Interestingly, the assumption on the mass ratio distribution seems to have a comparably strong impact, whereas the assumed eccentricity distribution has little effect on the completeness. Globally, we determine a completeness of $0.7 \pm 0.1$, however the completeness varies across the CMD. Our detection efficiency is highest for the SGB, RGB, and HB stars in the sample, whereas we appear to be most incomplete in detecting MS binaries, due to the differences in brightness between the individual groups, resulting in SNR ratio differences in the MUSE spectra and ultimately in differences in the achieved radial velocity accuracies. We note that due to the small sample size of the BSS and AGB populations, the detection efficiency for these groups is poorly constrained and so, as velocity error has the largest effect on detection efficiency, we adopt the efficiencies determined for MS and RGB stars, respectively, for these two groups. The full set of completeness values for each region of the CMD is listed in Tab.~\ref{table:Binary_Fraction} column (e).

        %In general, the recovery rates for the individual evolutionary stages follow the trends outlined by the overall sample. However, there are significant differences in terms of completeness across the CMD. 

        As discussed in Sec.~\ref{Subsec:Simulation_Setup}, we assume a constant velocity dispersion of $21~{\rm km\,s}^{-1}$ when generating the mock data set, though observations show that velocity dispersion decreases with radius \citep{Baumgardt_2018}. The velocity dispersion adopted has the most notable impact on the fraction of binary systems classified as ``soft'' (cf. Eq.~\ref{eqn:Binary_Hardness}), which are subsequently removed from the simulation. In order to understand the impact of the assumed velocity dispersion on our results, we use dispersion values of $21, 18, 15$ and $12~{\rm km\,s}^{-1}$, which represents the higher velocity dispersion within the central region of the cluster, within $\sim20\arcsec$ of the cluster centre, and the lower velocity dispersion at the half light radius of $\omega$~Cen, $\sim150\arcsec$ \citep[e.g.,][]{Baumgardt_2018}. We find that the fraction of recovered binaries decreases by $\sim0.05$ between $21$~and~$12~{\rm km\,s}^{-1}$, a decrease that is still within the uncertainties of the full sample completeness reported in column (e) of Tab.~\ref{table:Binary_Fraction}. We also test directly the assumptions of the cut value used to classify hard and soft binary systems. The default value for this is 1, which is in line with Eqn.~\ref{eqn:Binary_Hardness}. We vary the limit at which a system is classed as a hard binary, using cuts of 1, 0.4, 0.2 and 0. The latter three values effectively reduce the number of soft binaries to half, a quarter, and zero relative to the original number. We find that the fraction of recovered binaries decreases by a maximum of $\sim0.08$ when lowering the hard-soft binary limit. As the variation is still within the uncertainties reported in column (e) of Tab.~\ref{table:Binary_Fraction}, we are confident that the adopted threshold has no significant impact on our results.

    \subsection{Discovery fraction}
        \label{Subsec:Results_Disc_frac}

        With the reported photometric variables cleaned from the sample, the discovery fraction of $\omega$ Cen can be determined both globally and for different regions of the CMD, calculated by dividing the number of observed variable stars by the total number of stars. Having removed photometric variables from the sample, and by applying the minimum 6 epoch cut and using the probability calculation method discussed in Sec.~\ref{Subsec:Probability_calc} with a probability threshold $P_{\rm var}>0.8$, we find that out of the $19\,059$ stars in the sample, 275 show variability in their radial velocities. A breakdown of the number of stars and identified binary systems for each region of the CMD are shown in Tab.~\ref{table:Binary_Fraction} in columns (b) and (c), respectively. We find the global discovery fraction for $\omega$ Cen to be $1.4 \pm 0.1\%$, with the global discovery fraction and the fraction for the separate regions of the CMD shown in Tab.~\ref{table:Binary_Fraction} in column (d). This fraction varies depending on the region of the CMD that is considered. As discussed above, our efficiency is higher for brighter stars and at fixed brightness, also tends to increase for cooler stars. This being the case, we find that the discovery fraction amongst RGB stars is approximately $1\%$ higher than MS stars.

        Although the discovery fraction for most areas of the CMD is approximately 1-3\%, we note that the Blue Straggler Stars (BSS) have an enhanced discovery fraction of $13.6 \pm 5.1\%$. This is considerably higher than any of the other discovery fractions and suggests a $\sim 3\times10^{-5}$ probability that the discovery fraction of BSS is consistent with the global discovery fraction. This is, however, an expected result, as it is commonly assumed that BSSs form via binary interactions, like mergers or mass accretion from a binary companion \citep{Stryker_1993}.
    
    \subsection{Global binary fraction of Omega Centauri}
        \label{Subsec:Binary_Fraction}

        Using the discovery fractions presented in Sec.~\ref{Subsec:Results_Disc_frac} and the completeness corrections from Sec.~\ref{Subsec:Dependence_on_Orbital_Parameters}, we present the results for the binary fraction of $\omega$ Cen in Tab.~\ref{table:Binary_Fraction} column (f).
        
        \begin{table*}
            \centering
            \begin{tabular}{cccccc}
                \hline                               
                Stellar Type (a) & N$_{\rm stars}$(b) & N$_{\rm binaries}$ (c) & Discovery Fraction (d) & Sample Completeness (e) & Binary Fraction (f) \\ \hline
                Global fraction & 19059 & 275 & 1.4\% $\pm$ 0.1\% & 0.7 $\pm$ 0.1 & 2.1\% $\pm$ 0.4\%\\ %\\4.76\% $\pm$ 0.6\% \\
                MS & 14871 & 183 & 1.2\% $\pm$ 0.1\% & 0.6 $\pm$ 0.1 & 2.0\% $\pm$ 0.5\%\\ %4.92\% $\pm$ 0.6\% \\
                TO & 7897 & 84 & 1.1\% $\pm$ 0.1\% & 0.6 $\pm$ 0.1 & 1.6\% $\pm$ 0.4\%\\ %3.15\% $\pm$ 0.51\% \\
                SGB & 2396 & 23 & 1.0\% $\pm$ 0.3\% & 0.8 $\pm$ 0.1 & 1.2\% $\pm$ 0.4\%\\ %2.11\% $\pm$ 0.28\% \\
                RGB & 1811 & 41 & 2.3\% $\pm$ 0.4\% & 0.9 $\pm$ 0.1 & 2.4\% $\pm$ 0.4\%\\ %2.58\% $\pm$ 0.27\% \\
                HB & 232 & 6 & 2.6\% $\pm$ 1.3\% & 0.8 $\pm$ 0.1 & 3.0\% $\pm$ 1.5\%\\ %6.56\% $\pm$ 1.1\% \\
                AGB & 33 & 0 & 0\% & - & -\\ %3.12\% $\pm$ 0.58\% \\
                BSS & 44 & 6 & 13.6\% $\pm$ 5.1\% & 0.6 $\pm$ 0.1$^{*}$ & 21.9\% $\pm$ 9.5\%\\ 
                
                %17.2\% $\pm$ 4.07\\
                \hline

            \end{tabular}
        \caption{The binary fraction of $\omega$ Cen. Column (a) denotes the area of the CMD focused on for each calculation, along with the full sample. Columns (b) and (c) show the total number of stars and number of detected binary systems for each region of the CMD, respectively. Column (d) lists the discovery fraction of $\omega$ Cen as discussed in Sec.~\ref{Subsec:Results_Disc_frac}. Column (e) lists the completeness of each sample (see Sec.~\ref{Sec:Completeness_Correction}). Column (f) lists the fully corrected binary fraction of $\omega$ Cen. We see an increase in both the discovery and binary fractions particularly in the BSS region of the CMD.\\
        $^{*}$Using MS correction factor (see Sec.~\ref{Subsec:Dependence_on_Orbital_Parameters})}
        \label{table:Binary_Fraction}
        \end{table*}

        We report that the global corrected binary fraction of Omega Centauri is 2.1\% $\pm$ 0.4\%. This fraction is in reasonable agreement to the photometric estimate of $2.70\%\pm0.08\%$ obtained by \cite{Bellini_2017} and the upper limit of 5\% from \cite{Elson_1995}. Existing spectroscopic estimates of the binary fractions of globular clusters include the study of M4 by \cite{Sommariva_2009}, which reported a lower limit of $3\% \pm 0.3\%$, and the study of NGC 3201 by \cite{Giesers_2019}, which reported a fraction of $6.75\% \pm 0.72\%$. Both clusters are significantly less massive than $\omega$~Cen and \mbox{\citet{Milone_2012_photometry}} have shown that the binary fraction of Galactic globular clusters decreases with cluster luminosity (and hence mass). For the most massive clusters in their sample, \mbox{\citet{Milone_2012_photometry}} derive central binary fractions of few percent, in good agreement with our estimate for $\omega$~Cen.

        The results also show a larger binary fraction in the BSS population, with a binary fraction of $21.9\% \pm 9.5\%$. While the binary fraction has a large error associated due to the small sample size, there is a significantly higher percentage seen both in the binary fraction and the discovery fraction. A higher BSS binary fraction is also seen in the \cite{Giesers_2019} study of NGC 3201, at $57.5\% \pm 7.9\%$. This result is consistent with two formation scenarios: either mergers induced from interactions between binary systems and another binary or single star \citep[e.g.,][]{Leonard_1989,Fregeau_2004}, or from mass transfer within binary and triple systems \citep{Antonini_2016}.

        \subsubsection{Radial distribution of binaries}
            \label{Subsubsec:Radial_Distribution}

\begin{figure}
                \centering
                \includegraphics[scale=0.57]{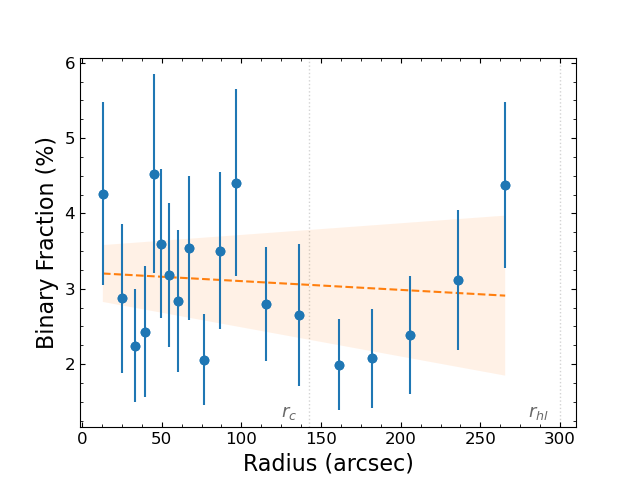}
                \caption{The binary fraction of $\omega$ Cen against the radius from the cluster centre. The data is binned such that one point represents 1000 stars. The best-fitting gradient of the distribution of $(-0.001 \pm 0.003)\,{\rm arcsec^{-1}}$ is shown as a dashed line, and the confidence interval as an orange-shaded area. The core radius and half light radius as reported by \citet{Harris_1996} are marked with dotted grey lines.}
                \label{fig:radial_distribution}
            \end{figure}

            To investigate the effects of mass segregation within the cluster, we determined the discovery fraction as a function of projected distance to the cluster centre, using the cluster centre position ($\alpha$, $\delta$)=(13h26m47.24s, -47\degr28\arcmin46.45\arcsec) as reported by \cite{Anderson_2010}. We separate the stars into bins of 1\,000 based on the distance from the cluster centre, and applied the completeness discussed in Sec.~\ref{Subsec:Dependence_on_Orbital_Parameters} to each bin. The results of this analysis are shown in Fig.~\ref{fig:radial_distribution}. Also shown in Fig.~\ref{fig:radial_distribution} is the best fitting linear gradient and its uncertainty interval.

            Through mass segregation, discussed in Sec.~\ref{Sec:Intro}, we would expect there to be some dependence on distance from the cluster centre, however, from Fig.~\ref{fig:radial_distribution} assuming a linear relation, we see approximately flat distribution, with a gradient of $(-0.001 \pm 0.003)\,{\rm arcsec^{-1}}$ inside the half mass radius of the cluster, which suggests that binary segregation in $\omega$~Cen is limited. This result agrees with results based on other tracers, such as a limited amount of energy equipartition \mbox{\citep{Watkins_2022}} or the shallow gradient in the slope of the mass function \mbox{\citep{Baumgardt_2023}}.
            
            Simulations of GCs by \cite{Aros_2021} suggest a flattened and lowered radial distribution of binary stars may also indicate an IMBH, however, a large collection of black holes may also have the same effect, albeit lessened. This is particularly important in light of the recent detection of high velocity stars at the centre of $\omega$~Cen by \citet{Haberle_2024}, which provides compelling evidence for the cluster to host an IMBH. The simulations by \citeauthor{Aros_2021} use simulated clusters of a lower mass than $\omega$ Cen with a binary fraction of 10\% and with binary properties specific to each cluster, which may make the results less applicable to the observed distribution. Hence, further investigations into the expected distribution of binary stars in $\omega$~Cen in either scenario are needed in order to draw firm conclusions from Fig.~\ref{fig:radial_distribution}.

            Although we initially assume a linear relation between radius and binary fraction, it is important to consider a non linear relation. The apparent dip seen in Fig.~\ref{fig:radial_distribution} between $\sim110\arcsec$ and $\sim220\arcsec$ shows similarities to the so called ``dynamical clock" presented by \cite{Ferraro_2012, Ferraro_2020}, which describes a link between the radial distribution of BSSs  and dynamical age. For dynamically young clusters in the early stages of mass segregation, BSSs from intermediate distances from the cluster centre migrate towards the centre of the cluster. Due to the longer relaxation time in the cluster outskirts, however, binaries in the outer regions of the cluster have not yet migrated into this intermediate region, resulting in a dip in the population of BSSs at this intermediate distance. The observations of the ``dynamical clock" has been corroborated by further observations of dynamically young clusters by \citet{Salinas_2012} and \citet{Dalessandro_2015}. Given that binary systems are impacted by mass segregation in a similar fashion as BSSs and that $\omega$~Cen is a dynamically young cluster, it is possible that the feature seen in the $110\arcsec$ and $220\arcsec$ region is an example of this relationship. In our analysis, we fit higher order polynomials to the data and, while higher order polynomials do improve the fit, the improvements seen were not statistically significant.

%        It also appears from the data that there is a lower fraction of RGB stars compared to the MS fraction. This suggests two explanations. The first is that there may be issues in assumptions made in creating the mock data set, however, much of the uncertainty in the parameter selection should be accounted for when testing, as discussed in Section \ref{Subsec:Sensitivity_to_orbital_parameters}. We do, however, accept that this is not a full model and simulation of omega cen and some accuracy may be lost.

%        The second, and more likely, explanation is connected to the velocity errors for areas of the CMD. RGB stars are more likely to have a lower velocity error, as their relative brightness allows for each star to be more well measured. This is in contrast to the MA branch, and therefore the global fraction, which typically has a higher average velocity error. This is reflected in the radial velocity data, as the mean velocity error for MS stars is $10.7$kms$_{-1}$, while the mean velocity error for RGB stars is $2.9$kms$_{-1}$. As the velocity error for RGB stars is lower, it makes us more sensitive to detecting binary systems that may be harder to detect, such as longer periods or low mass ratios.

\section{Conclusions}
    \label{Sec:Conclusions}

    In conclusion, using MUSE data, we have been able to determine the number of binary stars in $\omega$~Cen. We have then created mock data sets to rigorously test our assumptions of parameters and determine the completeness of our sample. We then combine these two elements to calculate a corrected binary fraction. In summary:
    
    %In this study, we have determined the number of binary stars in Omega Centauri, using the probability calculation method presented by \cite{Giesers_2019} with the data now available to us with MUSE, and combining this with a mock data set created to test the completeness of the sample. In summary:
    
    \begin{itemize}
        \item We have calculated that the global discovery fraction of $\omega$ Cen is $1.4 \pm 0.1\%$, calculated using the statistical method presented by \citet{Giesers_2019}. We limit our sample to stars with a minimum of 6 epochs, and only considering stars with a variability probability value of 0.8 in order to minimise false positives. In total, we find 275 likely binary systems from a sample of $19\,059$ stars. 
        \item We have created a mock set of data for $\omega$ Cen using randomly generated orbital parameters. The samples contained a set number of binary systems with known binary parameters in order to effectively test the completeness of the sample and calculate a corrected binary fraction of $\omega$ Cen.
        \item The completeness of the global discovery fraction was found to be 0.7 $\pm$ 0.1. This correction factor allows us to minimise the number of false positive detections to approximately $2\sigma$.
        \item By adjusting the discovery fraction with the completeness factor, we compute a global binary fraction of 2.1\% $\pm$ 0.4\%. This result agrees with the limit of $5\%$ from \cite{Elson_1995} and the photometric estimate derived in \mbox{\citet{Bellini_2017}}.
        \item From individual analysis of areas of the CMD, we find an increased binary fraction of stars in the BSS region of the CMD, at $21.9\% \pm 9.5\%$. The increased discovery and binary fraction on the BSS branch supports the theory that BSS evolution is linked to binary star evolution. 
        \item By investigating how the binary fraction of $\omega$ Cen evolves with radius, we find no radial trend, which is consistent with limited mass segregation \citep{Watkins_2022, Baumgardt_2023}
        \end{itemize}
    
    In upcoming studies focused on the binary population inside $\omega$~Cen, we will focus on the potential differences in the binary fractions of the multiple populations known to exist in the cluster and will present the results of Keplerian fits to the velocity curves obtained for the binaries identified in this paper.

\section*{Acknowledgements}

%\begin{itemize}
%    \item Marilyn matched the braga catalogue for me to the MUSE data 
%    \item Stefan provided the Kepler class
%    \item UKRI funding
%\end{itemize}
We thank the anonymous referee for their constructive feedback. F.W. acknowledges the joint studentship support from the Science and Technology Facilities Council and Liverpool John Moores University Faculty of Engineering and Technology (grant ST/V506874/1). S.K. acknowledges funding from UKRI in the form of a Future Leaders Fellowship (grant no. MR/T022868/1). S.S. acknowledges funding from STFC under the grant no. R276234. M.L. acknowledges funding from the Deutsche Forschungsgemeinschaft (grant LA 4383/4-1). We acknowledge the support from the Deutsche Forschungsgemeinschaft (DFG) through project DR~281/41-1 and from the German Ministry for Education and Science (BMBF Verbundforschung) through grants 05A14MGA, 05A17MGA, and 05A20MGA.

\section*{Data Availability}

%The inclusion of a Data Availability Statement is a requirement for articles published in MNRAS. Data Availability Statements provide a standardised format for readers to understand the availability of data underlying the research results described in the article. The statement may refer to original data generated in the course of the study or to third-party data analysed in the article. The statement should describe and provide means of access, where possible, by linking to the data or providing the required accession numbers for the relevant databases or DOIs.
%The data underlying this article will be shared on reasonable request to the corresponding author.

The raw MUSE data as well as reduced data cubes are available from the ESO archive. Mean velocity for the stars in our sample based on the same data have been published by \citet{Nitschai_2023}. The time-resolved velocities used in this study, including variability probabilities, will be published in the upcoming study by Saracino et al. (in prep) and are available from the authors upon request.

%refer to sara's upcoming paper, in report say that if the referee feels very strong, then this release can be brought forward,
%list of bin prob, ID and photo var



\bibliographystyle{mnras}
\bibliography{main} % if your bibtex file is called example.bib

%\begin{thebibliography}{99}
%\end{thebibliography}

%%%%%%%%%%%%%%%%%%%%%%%%%%%%%%%%%%%%%%%%%%%%%%%%%%

%%%%%%%%%%%%%%%%% APPENDICES %%%%%%%%%%%%%%%%%%%%%

%\appendix


%%%%%%%%%%%%%%%%%%%%%%%%%%%%%%%%%%%%%%%%%%%%%%%%%%

% Don't change these lines
\bsp	% typesetting comment
\label{lastpage}
\end{document}